\documentclass[10pt,twocolumn,prl,aps,floatfix,superscriptaddress,longbibliography]{revtex4-1}

\usepackage{color}
\usepackage{graphicx}
\usepackage{physics}
\usepackage{amsthm}
\usepackage{amsmath}
\usepackage{amssymb}
\usepackage{enumerate}
\usepackage{placeins}
\usepackage{booktabs}
\usepackage{dsfont}
\usepackage{yfonts}
\usepackage{hyperref}

\newcommand{\Eqref}[1]{Eq.~\eqref{#1}}

\usepackage[color=green!60]{todonotes}

\definecolor{red}{HTML}{FF595e}
\definecolor{orange}{HTML}{FE971A}
\definecolor{green}{HTML}{75D1B2}
\definecolor{blue}{HTML}{5380BA}
\definecolor{purple}{HTML}{6a4c93}

\AtBeginDocument{%
\heavyrulewidth=.08em
\lightrulewidth=.05em
\cmidrulewidth=.03em
\belowrulesep=.65ex
\belowbottomsep=0pt
\aboverulesep=.4ex
\abovetopsep=0pt
\cmidrulesep=\doublerulesep
\cmidrulekern=.5em
\defaultaddspace=.5em
}

\newif\ifarXiv
\arXivtrue 

\ifarXiv

\usepackage{pdfpages} 
\usepackage{pgffor} 
\usepackage{xr} 

\makeatletter
\AtBeginDocument{\let\LS@rot\@undefined}
\makeatother

\def\supplementfilename{supplement_unary_basis}

\pdfximage{\supplementfilename.pdf}
\def\numbersupplementpages{\the\pdflastximagepages}

\fi 

\begin{document}

\title{Balls and Walls: A Compact Unary Coding for Bosonic States}

\author{Hatem Barghathi}
\affiliation{Department of Physics and Astronomy, University of Tennessee, Knoxville, TN 37996, USA}

\author{Caleb Usadi}
\affiliation{Department of Physics, University of Vermont, Burlington, VT 05405, USA}

\author{Micah Beck}
\affiliation{Min H. Kao Department of Electrical Engineering and Computer Science, University of Tennessee, Knoxville, TN 37996, USA}

\author{Adrian Del Maestro}
\affiliation{Department of Physics and Astronomy, University of Tennessee, Knoxville, TN 37996, USA}
\affiliation{Min H. Kao Department of Electrical Engineering and Computer Science, University of Tennessee, Knoxville, TN 37996, USA}

\begin{abstract}
    We introduce a unary coding of bosonic occupation states based on the famous ``balls and walls'' counting for the number of configurations of $N$ indistinguishable particles on $L$ distinguishable sites.  Each state is represented by an integer with a human readable bit string that has a compositional structure allowing for the efficient application of operators that locally modify the number of bosons. By exploiting translational and inversion symmetries, we identify a speedup factor of order $L$ over current methods when generating the basis states of bosonic lattice models.  The unary coding is applied to a one-dimensional Bose-Hubbard Hamiltonian with up to $L=N=20$, and the time needed to generate the ground state block is reduced to a fraction of the diagonalization time. For the ground state symmetry resolved entanglement, we demonstrate that variational approaches restricting the local bosonic Hilbert space could result in an error that scales with system size.
\end{abstract}

\maketitle

\label{sec:introduction}
Finite lattice models represent an essential simplification in quantum many-body physics, where a set of discrete low energy degrees of freedom are sufficient to capture the relevant phases and phase transitions of a more complete high energy description.  The states of these quantum systems form a finite dimensional Hilbert space $\mathcal{H}$ and are thus amenable to a numerical representation (to arbitrary precision) through the exact diagonalization (ED) of an appropriate lattice Hamiltonian \cite{Lin:1990mw,Dagotto:1992tv,Lin:1993ed, Damski:2005,Weise:2008ed,Zhang:2010,Hu:2012,Sandvik:2010,Szabados:2012,Ravents:2017, Weinberg:2017,Weinberg:2019,westerhout:2021ls}. This provides a complete description of the eigenstates and related quantum many-body phenomena,  however, it is limited by the exponentially increasing cardinality  $\vert \mathcal{H}\vert \sim \mathrm{e}^L$ of $\mathcal{H}$ for $L$ lattice sites. 
This problem has motivated stochastic and variational approaches, such as quantum Monte Carlo (QMC) \cite{Blankenbecler:1981qt,Sandvik:1991pd,Prokofev:1998wl,Kozik:2010fv}, and the density matrix renormalization group (DMRG) \cite{White:1992xa,White:1993qd,Schollwock:2005,Schollwock:2011}. These methods are essential to our understanding of quantum many-body phenomena \cite{Kuhner:1998,Rapsch:1999,Kuhner:2000,Capogrosso:2007,LeBlanc:2015mj,Schafer:2021hm,Qin:2021}, however, ED still plays a crucial role in benchmarking novel methods, as well as in gaining access to the density matrix. While extremely efficient approaches exist for the Heisenberg model \cite{Lin:1990mw} and its SU($N$) generalizations \cite{Nataf:2014ar} with applications to fermionic lattice models of up to $\approx 40$ sites, the expanded Hilbert space of bosonic systems, allowing multiple occupancy of a single spatial mode, presents a formidable challenge \cite{Sundholm:2004un,Streltsov:2010xd,Szabados:2012,Ravents:2017}.

In this Letter we introduce a unary basis (UB) coding of bosonic occupation (Fock) states using a highly compact and compositional, yet still human-readable bit-string exploiting the famous \emph{balls and walls}  illustration of particles and sites used in Bose-Einstein counting. For example, for $N=L=11$:
\begin{align}
    \ket{\textcolor{red}{2},0,\textcolor{blue}{1},0,\textcolor{green}{3},0,\textcolor{orange}{1},0,0,0,\textcolor{purple}{4}} &\equiv\vert \textcolor{red}{\bullet\bullet} \vert ~ \vert \textcolor{blue}{\bullet}\vert ~ \vert\textcolor{green}{\bullet\bullet\bullet}\vert ~\vert\textcolor{orange}{\bullet}\vert ~\vert ~\vert ~\vert\textcolor{purple}{\bullet\bullet\bullet\bullet~~}\nonumber\\ I_{UB}=\texttt{2541296} &\equiv \texttt{1\textcolor{red}{00}11\textcolor{blue}{0}11\textcolor{green}{000}11\textcolor{orange}{0}1111\textcolor{purple}{0000}} 
\label{eq:UB}
\end{align}
where $\ket{n_1,n_2,\dots,n_L}$ is a characteristic basis state with $n_j$ the occupancy of site $j$. In the UB coding, \texttt{1}'s denote sites with the following number of \texttt{0}'s corresponding to that site's occupation, and each basis state can be represented as an integer in base 10. In addition to reducing the memory required to store the basis by a factor $L$, this approach provides rapid access to $n_j$ and significantly accelerates the action of local operators via bitwise operations on $I_{UB}$. An implementation in the presence of lattice symmetries yields a reduction of computational complexity by a factor of $L$ reducing the time needed to generate the basis for a one dimensional (1D) Bose-Hubbard (BH) model to a fraction of that needed to obtain the ground state via iteration. This yields a practical speedup of over $20 \times$ compared to current methods based on unique integer ordering of permanents \cite{Szabados:2012,Ravents:2017} for system sizes up to $L=N=20$.

We validate the utility of the unary basis by studying entanglement at the critical point of the 1D BH model at unit filling for both spatial mode and particle bipartitions as well as in the presence of a U$(1)$ symmetry fixing the total number of particles.  We show that in the latter case, the use of an ED method that retains the entire Hilbert space (as opposed to ``soft-spin'' approaches that fix the maximum occupancy on a site) is essential to obtain accurate results for the symmetry resolved entanglement.

In the remainder of the Letter, we demonstrate that close to unit filling, the unary basis is compact (fully utilizes the information stored in its bitwise representation) and describe how its compositional nature facilitates a machine and compiler dependent constant speedup factor for basis and Hamiltonian generation.  We then extend to the case of lattice symmetries where the unary basis achieves a reduction of algorithmic complexity.  After benchmarking on the BH model, we conclude with a discussion of potential future improvements including extensions to higher dimensions. To facilitate adoption, a software implementation is included in Ref.~\cite{repo}.

\noindent\emph{Background} -- For both fermionic and bosonic lattice models, it is natural to diagonalize any local $N$-particle Hamiltonian in a basis of spatial modes $\ket{n_1,n_2,\dots,n_L}$. For fermions, the Pauli exclusion principle restricts $n_j = 0,1$ and each of the $\lvert \mathcal{H}_f \rvert = \binom{L}{N}$ basis states can be encoded as a binary word of length $L$ that can be stored as a $2^{\lceil \log_2 L \rceil}$-bit integer ($\lceil \dots \rceil$ is the integer ceiling function) as is implemented in commonly used ED software \cite{Bauer:2011nd,Kawamura:2017so,Weinberg:2017,Weinberg:2019}. For bosons, the possibility of multiple occupancy, $n_j=0,\dots,N$ on any site enlarges the Hilbert space to $\abs{\mathcal{H}} =\binom{N+L-1}{N}$ with basis vectors for the occupation states naturally parameterized by $L\vert \mathcal{H}\vert$ integers.  The memory required to store the Hamiltonian of such a system has an upper bound $\sim\abs{\mathcal{H}}^2$ and thus dwarfs that needed to store the basis states. However, in bosonic lattice models with limited range hopping, the Hamiltonian matrix is sparse, reducing its storage cost to $\sim L\vert \mathcal{H}\vert$ and encoding the basis as arrays of site occupations now composes a leading share of the required memory.  To address this problem,
permanent ordering (PO) schemes \cite{Sundholm:2004un,Streltsov:2010xd,Szabados:2012,Ravents:2017} have been introduced which assign a unique contiguous integer label $I_{PO}(n_1,\dots,n_L)$ to each occupation state via an iterative procedure of complexity $\mathrm{O}(L)$. To further avoid the memory impact of storing $\qty{I_{PO}}$, a lookup table combined with an appropriate $\mathrm{O}(L)$ inverse function $I_{PO}^{-1}$ \cite{Ravents:2017}, can be implemented to gain access to site occupations numbers $n_j$. While this method has proven to be extremely effective, for example, in studying many-body localization in bosonic systems \cite{Sierant:2017jh,Hopjan:2020do} it has yet to be extended to take advantage of lattice symmetries (e.g.\@ translation, inversion) where larger system sizes can be studied as the maximum number of non-zero Hamiltonian matrix elements is reduced to $\sim \abs{\mathcal{H}}$. Here it is crucially important to suppress the now leading order memory share ($\sim L\vert \mathcal{H}\vert$) of the basis.

\noindent\emph{Unary Basis} -- In order to study the largest possible systems, we introduce an integer labelling scheme motivated by the formal equivalence between occupation states of $N$ bosons on $L$ sites, and those of $N$ fermions on $N+L$ lattice sites \cite{Streltsov:2010xd}. By reinterpreting bits corresponding to the presence (\texttt{1}) or absence (\texttt{0}) of a fermion with a site boundary and boson respectively, the resulting set of non-contiguous $\abs{\mathcal{H}}$ integers  $\qty{I_{UB}}$ can encode information about the occupation state directly in their bit string (see Eq.~\eqref{eq:UB}) while remaining as efficient as a binary encoding for fermions. This can be quantified by defining 
the efficiency of the representation as $\eta = \log_2\abs{\mathcal{H}}/\log_2 2^{L+N}$, 
providing a maximum efficiency at unit filling ($\nu=N/L=1$, with $N$ and $ L\gg 1$) of $\eta=1-(1/4L)\log_2 L+ \mathrm{O}(L^{-1})$, where for $L=20$, $\eta \simeq 0.925$. 
Comparing with an alternative approach where $N$ bosons are mapped to a spin model with $S=N/2$ \cite{Weinberg:2017,Weinberg:2019}, that uses a sequence of $L\lceil\log_2(N+1)\rceil$ bits coding $n_j \le N$ on each site,  the corresponding efficiency at unit filling is  $\eta \approx2/\lceil\log_2(L)\rceil-1/(2L)+ \mathrm{O}((L\log_2L)^{-1})$ which is smaller by a factor of  $2/\lceil\log_2(L)\rceil$ compared to the unary coding, and for $L=20$, $\eta \approx 0.37$.

In addition to memory compactness, the UB coding can accelerate the inverse operation $I_{UB}^{-1}$ to obtain access to occupation numbers. 
Given a 64-bit integer $I_{UB}$ representing $\ket{n_1,\dots,n_L}$, the number of particles on the first site $n_1$ can be found by counting the trailing zeros in the bit string of $I_{UB}$ (a ``compiler builtin''). Shifting the bits of $I_{UB}$ by $n_1+1$ allows for reading the number of particles on the next site, and by repeating these bitwise operations $L$ times, the corresponding occupation vector (OV) $\ket{n_1,\dots,n_L}$ can be constructed. This procedure can be further sped up by viewing the $64$ bits of $I_{UB}$ as a sequence of four $16$-bit integers from which the corresponding number of sites and occupation numbers can be obtained by direct lookup of the $2^{16}$ possibilities then recomposed to generate the corresponding OV.  
We find that generating OVs for $\qty{I_{UB}}$ is more than $4\times$ faster than obtaining them on the fly using permanent ordering, at a cost of additional memory usage corresponding to only $1/L$ of that needed to store the system Hamiltonian in the absence of translational symmetry. 

\noindent\emph{Lattice Symmetries} -- The utility of the unary coding can be extended to treat systems that preserve Hamiltonian symmetries such as translation and inversion (see Supplemental Material \cite{supplemental}). 
If the Hamiltonian commutes with the translation operator $\widehat{T}$,  it has a block diagonal structure where each of the $L$ blocks has a quasi-momentum index $q$ and contains a maximum number of non-zero elements that scales as $\sim \vert \mathcal{H}\vert$ (for short range hopping).  The resulting number of translationally symmetrized basis states (the $q^{th}$ degenerate set of the eigenstates of $\widehat{T}$) scales as $\sim \vert \mathcal{H}\vert/L$.  For a given $q$ and OV,
an eigenvector $\ket{\phi_{\alpha,q}}$ of $\widehat{T}$ can be generated: $\ket{\phi_{\alpha,q}}={L}_{\alpha}^{-1/2} \sum_{j=0}^{{L}_{\alpha}-1}\exp\left[-2\imath \pi j q/{L}_{\alpha}\right]\widehat{T}^j\vert n_1, n_2,\dots,n_L\rangle$, where $\alpha$ is the index of a \emph{cycle} with length ${L}_{\alpha}\leq L$ \footnote{${L}_{\alpha}$ is the minimum number of cyclic shifts which maps a given occupation vector onto itself. Except for an exponentially small fraction of states, ${L}_{\alpha}=L$ (see supplemental material \cite{supplemental}).} generated by repeatedly acting on a given $\vert n_1, n_2,\dots,n_L\rangle$ with $\widehat{T}$.  Each cycle can be mapped to a set of eigenvectors of $\widehat{T}$ with the same number of elements.

Calculating matrix elements or expectation values of local operators requires access to all OVs in $\vert\phi_{q,\alpha}\rangle$, as well as the ability to reconstruct the state $\vert\phi_{q,\alpha}\rangle$ where a given OV appears. The naive storage cost for direct lookup of this information is at least double that needed for $\qty{I_{UB}}$.  Alternatively, one can trade memory with computational complexity by storing ${L}_{\alpha}$ for each cycle, as well as an extremal integer $I_\alpha$ that represents one OV characteristic of the cycle, (i.e. $I_{PO}$ or its unary coding $I_{UB}$).  Thus, when inverting $I_\alpha$ to an OV, the rest of the cycle can be obtained by ${L}_{\alpha}$ translations. If $I_\alpha \in \qty{I_{UB}}$ the unary coding of the OVs of the cycle can be obtained by bit shifting $I_\alpha$. As a single 64-bit instruction can shift a vector of 8 bytes (using parallel intra-processor bit paths) faster than moving each byte sequentially (using multiple memory operations), the UB coding provides a reliable constant speedup.

The inverse process of finding the cycle $\alpha$ where a given $\ket{n_1,\dots,n_L}$ appears can be done in four steps: (i) cyclic shifting $L$ times; (ii) converting each of the $L$ OVs to their integer representation; (iii) finding the characteristic integer; and (iv) performing a fast search for $\alpha$ in the ordered list of integers $I_\alpha$.  While steps (i) and (iii)-(iv) require at most  $\mathrm{O}(L)$ operations, the time complexity of step (ii) is $\mathrm{O}(L^2)$, as it consists of repeating the operation of converting an OV to an integer $L$ times. However, by exploiting the unary coding, the asymptotically slowest step (ii) is not required, as the shifting process (i) can be directly applied to $I_{UB}$ to obtain the targeted list of integers. This represents a significant opportunity for a reduction in complexity and speedup over PO methods.

%

\noindent \emph{Benchmarking} -- As a physically relevant test case, we consider the unary coding for $N$ bosons on a ring of $L=N$ sites ($\nu=1$) governed by the Bose-Hubbard Hamiltonian:
\begin{equation}
    \widehat{H}=-\sum_{i=1}^{L} \qty( b_{i+1}^\dagger b_{i}^{\phantom{\dag}} +\text{h.c.})+ \frac{U}{2}\sum_{i=1}^{L} n_i(n_i - 1)
\label{eq:Bose-Hubbard},
\end{equation}
where $n_i=b_i^\dagger b^{\phantom{\dagger}}_i$, $[b^{\phantom{\dagger}}_i,b_j^\dagger]=\delta_{i,j}$ and $b_{L+1} = b_1$.
The on site interaction $U>0$ is measured in units of the hopping. 
Eq.~\eqref{eq:Bose-Hubbard} exhibits a continuous phase transition from a superfluid ($U\ll 1$) to an insulator ($U\gg 1$) at $U \approx 3.3$ \cite{Kuhner:1998,Carrasquilla:2013,Boris:2016qt}. Using the unary coding of the basis, as well as translational ($q=0,\dots,L-1$) and inversion ($r=\pm 1$) symmetries on a ring,  we have obtained the ground state $\ket{\Psi_0}$ of \Eqref{eq:Bose-Hubbard} via exact diagonalization for up to $N=20$ bosons on $L=20$ sites.

\begin{figure}[t!]
\begin{center}
\includegraphics[width=1.0\columnwidth]{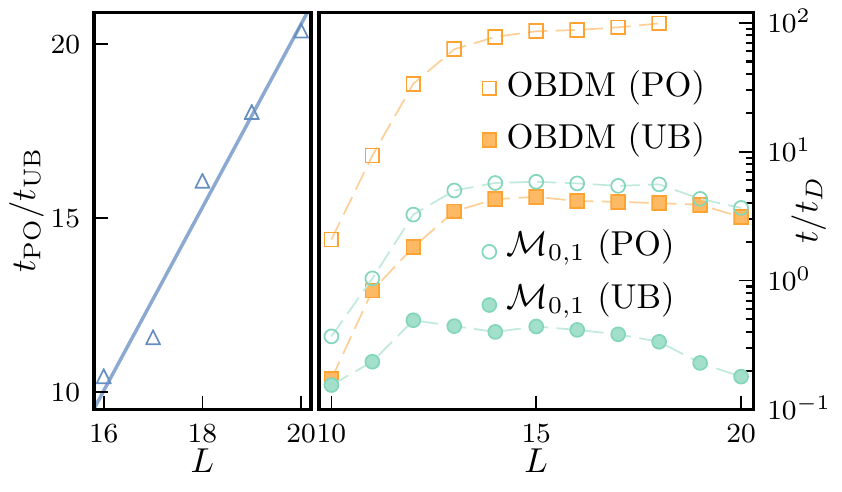} 
\end{center}
\caption{Computational efficiency of the unary basis (UB) and permanent ordering (PO) encodings. 
    Left: The ratio $t_{\rm{PO}}/t_{\rm{UB}}$ of wall clock times needed to construct the ground state block using the permanent ordering (PO) and unary basis (UB) codings as a function of system size $L$. The line is a linear fit demonstrating the complexity reduction discussed in the text.  
Right: The wall clock time ratio $t/t_D$ vs. system size $L$, where $t_D$ is the time consumed in diagonalizing the ground state block $\mathcal{M}_{q=0,r=1}$ of the Bose-Hubbard Hamiltonian and $t$ represents the time spent in constructing $\mathcal{M}_{q=0,r=1}$ (circles) or calculating the elements of the one body density matrix (OBDM) in the ground state (squares). 
} 
\label{fig:TimeScaling}
\end{figure}

To compare the efficiency of the PO and UB methods, we measure the times $t_{\rm{PO}}$ and $t_{\rm{UB}}$ spent in constructing the ground state block $\mathcal{M}_{q=0,r=1}$ of the sparse matrix representing $\widehat{H}$. We find that the ratio $t_{\rm{PO}}/t_{\rm{UB}}$ (Fig.~\ref{fig:TimeScaling} left), scales as $\mathrm{O}(L)$, as expected from steps (i)--(iv) described above. 
For $L=N=20$, we find a speedup of $t_{\rm{PO}}/t_{\rm{UB}} \gtrapprox 20$. The practicality of this linear in $L$ speedup becomes evident when measuring in units of $t_D$, the time required to iteratively obtain the ground state $\ket{\Psi_0}$ from $\mathcal{M}_{q=0,r=1}$.  Again, for $L=N=20$, we find  $t_{BU}/t_D \approx 0.19$ compared to  $t_{PO}/t_D \approx 3.96$, as illustrated in the right panel of Fig.~\ref{fig:TimeScaling}. 
Thus, employing the unary coding can practically reduce computation time by a factor of $(t_{PO}+t_D )/(t_{UB}+t_D )\gtrapprox 4$.  An even larger larger computational speed up ($\gtrapprox 20$ for $L=N=18$) is identified when calculating the elements of the one-body density matrix (OBDM) $\langle b_i^\dagger b^{\phantom{\dagger}}_j\rangle$ (right panel of Fig.~\ref{fig:TimeScaling}).

\noindent \emph{Application to Entanglement} -- As a further illustration of the utility of the unary coding, including the importance of using an unrestricted local bosonic Hilbert space, we investigate several measures of entanglement in $\ket{\Psi_0}$.  In all calculations, the required memory was less than $1.5$ terabyte (TB).

\begin{figure}[t]
\begin{center}
\includegraphics[width=1.0\columnwidth]{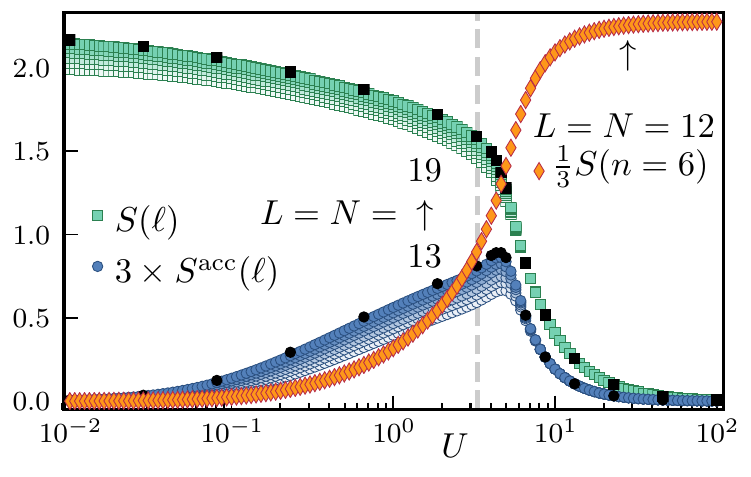} 
\end{center}
\caption{Mode entanglement $S(\ell)$ and $S^{\rm{acc}}(\ell)$) and particle entanglement entropy $S(n)$ in the ground state of the BH model as a function of interaction strength $U$ at unit filling ($L=N$). For mode entanglement, the partition size is fixed at $\ell=\lceil L/2\rceil$, where $f(x)=\lceil x\rceil$ is the least integer function} 
\label{fig:EntanglementMeasures}
\end{figure}

Starting with mode entanglement, by partitioning the spatial modes describing the pure state $\rho = \ket{\Psi_0}\bra{\Psi_0}$ into  $\ell$ consecutive sites $A_{\ell}$ and its complement $\bar{A}_{\ell}$, corresponding to the remaining $L-\ell$ modes, we obtain the reduced density matrix $\rho_{A_{\ell}} = \Tr_{\bar{A}_{\ell}}\, \rho$ by tracing out all degrees of freedom in $\bar{A}_{\ell}$. The von Neumann entanglement entropy $S(\ell) = -\Tr \rho_{A_{\ell}} \ln \rho_{A_{\ell}}$ quantifying the amount of entanglement that exists between $A_{\ell}$ and $\bar{A}_{\ell}$ \cite{Horodecki:2009gb}  
is shown as square symbols in Fig.~\ref{fig:EntanglementMeasures}.  The results demonstrate the reduction of spatial mode entanglement as the on-site repulsion $U$ is increased and the system transitions from a superfluid to localized phase. 

With access to an exact representation of the ground state $\ket{\Psi_0}$, we can also study particle entanglement \cite{Haque:2007il,Zozulya:2007jw,Zozulya:2008kb, Haque:2009df, Liu:2010pe, Herdman:2014jq, Herdman:2014ey, Herdman:2015gx, Rammelmuller:2017om, Barghathi:2017fv} where the bipartition of the Hilbert space is performed over particle labels (i.e.\@ the $N$ particles of the system are partitioned by distinguishing a set of $n$ particles from the remaining $N-n$ particles). This is most easily performed using the first quantized basis $\vert i_1,\dots, i_N\rangle$, where $i_k \in \qty{1,\dots,L}$ is the lattice coordinate of the particle labeled $k \in \qty{1,\dots,N}$. The familiar $n$-particle reduced density matrix of partition ${A_{n}}$ is then:

\begin{align*}
    \rho_{A_{n}}^{\qty{i_k}_1^n,\qty{j_k}_1^n}=\!\!\sum_{\qty{i_k}_{n+1}^{N}}\!\!\!\! \Psi^{\dag}(\qty{i_k}_1^n, \qty{i_k}_{n+1}^{N})\Psi(\qty{j_k}_1^n,\qty{i_k}_{n+1}^{N}),
\end{align*}
where $\qty{i_k}_a^b \equiv \qty{i_a,i_{a+1},\dots,i_{b-1},i_b}$. 
The $n$-particle entanglement entropy is $S(n) = -\Tr \rho_{A_{n}} \ln \rho_{A_{n}}$ with results shown as diamonds in Fig.~\ref{fig:EntanglementMeasures}. 
In contrast to $S(\ell)$, $S(n)$ is large in the insulating phase, vanishes as $U\to 0$ and is extremely difficult to calculate via DMRG for $n>2$ \cite{Kurashige:2014ca}. 

The interplay between fixed particle number and fluctuations between spatial modes, as reflected in symmetry resolved entanglement, is a subject of growing interest \cite{Murciano:2020,TanRyu:2020,Capizzi:2020,FraenkelGoldstein:2020,FeldmanGoldstein:2019,BarghathiCasiano-DiazDelMaestro:2019,BonsignoriRuggieroCalabrese:2019,Goldstein:2018kf,KieferEmmanouilidisUnanyanFleischhauer:2020,MurcianoDiGiulioCalabrese:2020,MurcianoDiGiulioCalabrese:2020FT,Benatti:2020,Turkeshi:2020,Faiez:2020,Horvath:2020,Bonsignori:2020,deGroot:2020,zhao:2021arXiv,fraenkel:2021arXiv,Horvath:2021,Parez:2021,Capizzi:2021arXiv,Murciano:2021,Estienne:2121,Horvath:2021arXiv,BarghathiHerdmanDelMaestro:2018,Melko:2016bo}. Here, the presence of conservation laws (e.g.\@ fixed $N$), reduces the amount of entanglement that is accessible for quantum information processing \cite{Bartlett:2003ud,WisemanVaccaro:2003,WisemanBartlettVaccaro:2004,VaccaroAnselmiWiseman:2003,SchuchVerstraeteCirac:2004,DunninghamRauBurnett:2005,CramerPlenioWunderlich:2011,KlichLevitov:2008}. The accessible entanglement can be quantified as $S^{\rm{acc}}(\ell)=\sum_{n=0}^{N} P_n S(\ell;n)$, where $S(\ell;n)=-\Tr \rho_{A_{\ell},n} \ln \rho_{A_{\ell},n}$ is the von Neumann entanglement of $\rho_{A_{\ell},n} = {\widehat{\mathcal{P}}}_{A_{\ell},n} \rho_{A_\ell} {\widehat{\mathcal{P}}}_{A_{\ell},n}/P_n$ obtained via projection by ${\widehat{\mathcal{P}}}_{A_{\ell},n}$ that fixes the number of particles in $A_{\ell}$ to $n$, and due to the conservation of the total number of particles $N$, also fixes the number of particles in $\bar{A}_{\ell}$ to $N-n$.  Here  $P_n = \mathrm{Tr}\, {\widehat{\mathcal{P}}}_{A_\ell,n} \rho_{A_\ell}{\mathcal{P}}_{A_\ell,n}$ is the probability of finding $n$ particles in $A_{\ell}$. Figure~\ref{fig:EntanglementMeasures} demonstrates that the accessible entanglement entropy vanishes at the extremes of the two competing phases, and peaks near the critical point. 

\begin{figure}[t]
\begin{center}
\includegraphics[width=1.0\columnwidth]{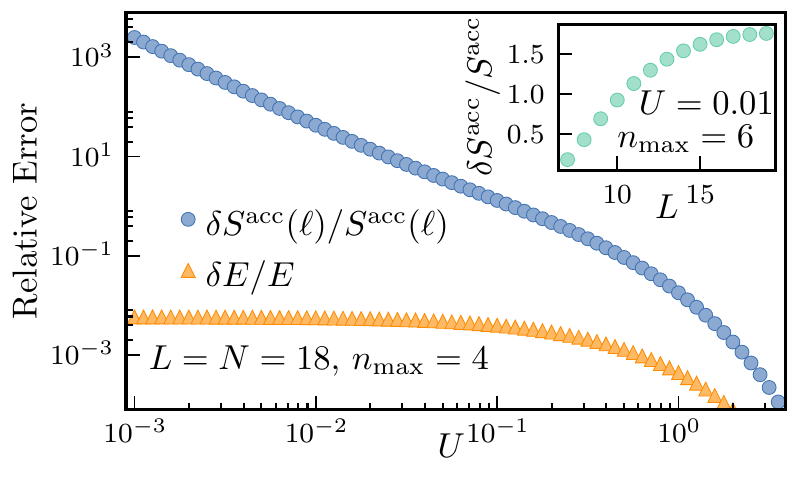} 
\end{center}
\caption{The relative error in the accessible entanglement $S^{\rm{acc}}(\ell=9)$ and the energy $E$ of the BH ground state,  arising from fixing the maximum occupancy at all sites to $n_{\rm{max}}=4$,  as a function of interaction strength $U/t$ at $L=N=18$. Inset: Scaling of the relative error $\delta S^{\rm{acc}}(\ell)/S^{\rm{acc}}(\ell)$ with the system size $L$ at $U =0.01$ and $n_{\rm{max}}=6$, where $\ell=\lceil L/2\rceil$.} 
\label{fig:RestrictedOccupationNumber}
\end{figure}

In general, restricting the local Hilbert space by enforcing an upper limit $n_{\rm{max}}$ on the occupation number per site is a widely used approximation when performing the ED of a bosonic Hamiltonian, or the DMRG representation of its ground state \cite{lauchli:2013,Carrasquilla:2013}.  The validity of this approximation is often judged based on the convergence of observables, (e.g.\@ the ground state energy, the average occupation per site, or its fluctuations \cite{Szabados:2012}) with $n_{\rm max}$.  However, the truncated degrees of freedom could contribute significantly to other observables that crucially depend on particle number fluctuations.  To illustrate this in the BH model, we calculate the error arising from restricting $n_{\rm{max}}=4$, in both the ground state energy $E$ and $S^{\rm{acc}}(\ell)$. Figure~\ref{fig:RestrictedOccupationNumber} shows that even though the error in $E$ is less than $1\%$,  the relative error in $S^{\rm{acc}}(\ell)$ could be three orders of magnitude larger in the superfluid phase. 
%
Moreover, the relative error increases with $L$ as demonstrated in the inset of Fig.~\ref{fig:RestrictedOccupationNumber}. To intuitively understand the origin of such an error scaling with $L$, we consider the simpler case of the bosonic density reduced to half-filling. In this case, $S^{\rm{acc}}$ still vanishes as $U\to0$. However, if we now enforce the extreme constraint of $n_{\rm{max}}=1$, we end up with a hardcore bosons, which for an appropriate parity of $N$, possesses the same $S^{\rm{acc}}(\ell=L/2)$ for a partition of consecutive sites as a system of non-interacting fermions.  Here, it is known that $S^{\rm{acc}}\sim \ln L$,  and thus, the error will have a similar scaling \cite{KlichLevitov:2008,BarghathiHerdmanDelMaestro:2018}.

\noindent\textit{Discussion} --
In this paper we have introduced a compact unary coding for bosonic basis states that is both human and machine readable by exploiting the trick of enumerating configurations of indistinguishable particles on distinguishable sites through combinatorial counting of \emph{balls and walls}.  Its implementation in the presence of lattice symmetries reduces both the computational complexity of generation, and the memory needed to store occupation vectors $\ket{n_1,\dots,n_L}$, by a factor of the system size $L$ over currently utilized methods.  This allowed us to study up to $N=20$ bosons at unit filling, achieving equivalence for bosonic lattice Hamiltonians with state of the art exact diagonalization methods for fermions ($L \approx 40$).   While still limited to finite bosonic clusters, having access to an efficient and compact exact numerical representation of the ground state allows for the study of information quantities such as the particle partition or symmetry resolved entanglement that are not amenable to measurement via variational (due to the difficulty in computing $n$-particle density matrices, or errors imposed by restricting the local Hilbert space), or stochastic (currently limited to R\'enyi entropies \cite{Hastings:2010dc}) algorithms. While we have focused a one-dimensional Bose-Hubbard Hamiltonian, the unary coding can be generally applied to any $D$-dimensional bosonic lattice model of up to $20$ sites utilizing standard large-memory resources. Significant prefactor speedups may be obtainable through union data types that allow for the simultaneous view of an occupation vector as an array of smaller integers.  We hope that the balls and walls coding will be rapidly incorporated into available exact diagonalization software, accelerating future studies of bosonic lattice Hamiltonians.

\acknowledgements
This work was supported in part by the NSF under Grant No.~DMR-2041995.


\FloatBarrier

\bibliography{refs}

\ifarXiv
    

\section*{Supplementary material}

# Supplemental Material for: "Balls and Walls: A Compact Unary Coding for Bosonic States"

Hatem Barghathi, Caleb Usadi, Micah Beck, and Adrian Del Maestro

In this supplement we include additional details on the application of lattice symmetries leading to the timing results presented in the main text. All computations were performed at the Infrastructure for Scientific Applications and Advanced Computing (ISAAC) at the University of Tennessee, Knoxville on a Intel(R) Xeon(R) Gold 6248R CPU 3.00 GHz node with 1536 GB of memory. In order to obtain accurate method-to-method comparisons, calculations were restricted to a single thread. While further decreases in wall clock time could be obtained using shared or distributed memory parallelism, relative timings as displayed in Figure 1 of the main text would be mostly unaffected.

## TRANSLATIONAL AND INVERSION SYMMETRIES

Here we discuss the spectra and eigenstates of the translation ($\widehat{T}$) and reflection ($\widehat{R}$) operators in the basis of bosonic occupation vectors and provide further details on how employing such symmetries in lattice models simplify the structure of the matrix representation of the governing Hamiltonian $\widehat{H}$.

### Translational Symmetry

The translation (shift) operator $\widehat{T}$ is defined by $\widehat{T} b_i \widehat{T}^\dagger = b_{i+1}$, where, $b_{i+L} = b_i$ as dictated by the periodic boundary conditions. Here, $[b_i, b_j^\dagger] = \delta_{i,j}$, where, $b_i$ and $b_i^\dagger$ are bosonic annihilation and creation operators at site $i$, respectively. From the definition of the unitary operator $\widehat{T}$, $\widehat{T}^\dagger b_i \widehat{T} = b_{i-1}$ and $\widehat{T}^L = \mathbb{1}$ (where $\mathbb{1}$ is the identity operator) resulting in a degenerate spectra with $L$ eigenvalues: $\exp[2\imath\pi q/L]$, where $q = 0, 1, \ldots, L-1$.

If $\widehat{H}$ is translationally invariant, i.e., the commutation relation $[\widehat{H}, \widehat{T}] = 0$, then $\widehat{H}$ can be decomposed as a block diagonal matrix, where each of the $L$ blocks $\mathcal{M}_q$ corresponds to one of the $q$ values. By repeatedly acting on a given occupation vector $|n_1, n_2, \ldots, n_L\rangle$ with $\widehat{T}$, a number of $L_\alpha \leq L$ independent occupation vectors, including the original one, can be generated. We call this a *cycle* and $\alpha$ is the cycle-index, i.e., $\widehat{T}^{L_\alpha}|n_1, n_2, \ldots, n_L\rangle = |n_1, n_2, \ldots, n_L\rangle$. If the occupation vector $|n_1, n_2, \ldots, n_L\rangle$ contains a fixed pattern of bosons repeated $m_\alpha$ times, then the number of elements in the corresponding cycle, or the cycle length, will be $L_\alpha = L/m_\alpha$. Obviously, $m_\alpha$ must be a common divisor of $L$ and $N$.

As a specific example, , consider $L = N = 12$, then the set of common divisors of $L$ and $N$ is $\{m_\alpha\} = \{1, 2, 3, 4, 6, 12\}$, e.g., the occupation vectors $|1, 2, 0, 1, 1, 1, 1, 2, 0, 1, 1, 1\rangle$ belongs to a cycle of with $L/2$ elements, while $|2, 0, 1, 1, 2, 0, 1, 1, 2, 0, 1, 1\rangle$ belongs to a cycle of with $L/3$ elements. Each cycle contains $L_\alpha$ eigenvectors of $\widehat{T}$:

$$|\phi_{\alpha,q}\rangle = \frac{1}{\sqrt{L_\alpha}} \sum_{j=0}^{L_\alpha-1} e^{-2\imath\pi j q_\alpha/L_\alpha} \widehat{T}^j |n_1, n_2, \ldots, n_L\rangle, \tag{S1}$$

with $q_\alpha = 0, 1, \ldots, L_\alpha - 1$. Here, $q = m_\alpha q_\alpha$. For a system with $N$ bosons on $L$ sites, if we denote the number of cycles with length $L$ as $\mathsf{Cyc}(L, N)$ then the number of cycles with length $L_\alpha = L/m_\alpha$ is $\mathsf{Cyc}(L/m_\alpha, N/m_\alpha)$ and thus the total number of cycles is:

$$\mathsf{Cyc}_{\mathrm{tot}}(L, N) = \sum_{\{m_\alpha\}} \mathsf{Cyc}\left(\frac{L}{m_\alpha}, \frac{N}{m_\alpha}\right). \tag{S2}$$

In this case, The total number of bosonic occupation vectors $\left(\!\binom{L}{N}\!\right) = \binom{L+N-1}{N}$ can be obtained from the number of cycles as

$$\left(\!\binom{L}{N}\!\right) = \sum_{\{m_\alpha\}} \mathsf{Cyc}\left(\frac{L}{m_\alpha}, \frac{N}{m_\alpha}\right) \frac{L}{m_\alpha}, \tag{S3}$$

and solving for $\mathsf{Cyc}(L, N)$, we obtain

$$\mathsf{Cyc}(L, N) = \left(\!\binom{L}{N}\!\right) \frac{1}{L} - \sum_{\{m_\alpha\}\setminus\{1\}} \mathsf{Cyc}\left(\frac{L}{m_\alpha}, \frac{N}{m_\alpha}\right) \frac{1}{m_\alpha}. \tag{S4}$$

Eq. (S4) can then be applied recursively to obtain $\mathsf{Cyc}\left(L_\alpha = \frac{L}{m_\alpha}, N_\alpha = \frac{N}{m_\alpha}\right)$ for each $\alpha$, where

$$\mathsf{Cyc}(L_\alpha, N_\alpha) = \left(\!\!\binom{L_\alpha}{N_\alpha}\!\!\right) \frac{1}{L_\alpha} - \sum_{\{\tilde{m}_{\tilde{\alpha}}\}\backslash\{1\}} \mathsf{Cyc}\left(\frac{L_\alpha}{\tilde{m}_{\tilde{\alpha}}}, \frac{N_\alpha}{\tilde{m}_{\tilde{\alpha}}}\right)\frac{1}{\tilde{m}_{\tilde{\alpha}}}, \tag{S5}$$

and $\{\tilde{m}_{\tilde{\alpha}}\}$ are the common divisors of $L_\alpha$ and $N_\alpha$.

In the special case where $N$ and $L$ are relatively prime, then all cycles will have length $L$ and thus their number is $\left(\!\!\binom{L}{N}\!\!\right)\frac{1}{L}$.

By examining Eq. (S4), we see that $\mathsf{Cyc}(L,N)$ deviates from $\left(\!\!\binom{L}{N}\!\!\right)\frac{1}{L}$ by $\sum_{\{m_\alpha\}\backslash\{1\}}\mathsf{Cyc}\left(\frac{L}{m_\alpha},\frac{N}{m_\alpha}\right)\frac{1}{m_\alpha}$. Equation (S5) then dictates that

$$\sum_{\{m_\alpha\}\backslash\{1\}} \mathsf{Cyc}\left(\frac{L}{m_\alpha}, \frac{N}{m_\alpha}\right)\frac{1}{m_\alpha} \leq \frac{1}{L}\sum_{\{m_\alpha\}\backslash\{1\}}\left(\!\!\binom{L_\alpha}{N_\alpha}\!\!\right). \tag{S6}$$

Given that $2 \leq m_\alpha \leq \min(L,N) \leq L$, we find

$$\sum_{\{m_\alpha\}\backslash\{1\}} \mathsf{Cyc}\left(\frac{L}{m_\alpha}, \frac{N}{m_\alpha}\right)\frac{1}{m_\alpha} \leq \frac{1}{L}\left(\!\!\binom{\lceil L/2\rceil}{\lceil N/2\rceil}\!\!\right)\sum_{\{m_\alpha\}\backslash\{1\}} 1 < \left(\!\!\binom{\lceil L/2\rceil}{\lceil N/2\rceil}\!\!\right)\frac{\sigma_0(\min(L,N))}{L}, \tag{S7}$$

where $f(x) = \lceil x \rceil$ is the least integer function. Here $\sum_{\{m_\alpha\}} 1$ counts the number of common divisors of $L$ and $N$, which is bounded by $\sigma_0(\min(L,N))$, where $\sigma_0(n)$ is the number of divisors of the positive integer $n$. Similarly, we can write

$$\sum_{\{m_\alpha\}\backslash\{1\}} \mathsf{Cyc}\left(\frac{L}{m_\alpha}, \frac{N}{m_\alpha}\right) < \left(\!\!\binom{\lceil L/2\rceil}{\lceil N/2\rceil}\!\!\right)\frac{\sigma_1(\min(L,N))}{L}, \tag{S8}$$

where $\sigma_1(n)$ is the sum of the divisors of the positive integer $n$. Using Eqs. (S4) and (S7) we obtain

$$\mathsf{Cyc}(L,N) > \frac{1}{L}\left(\!\!\binom{L}{N}\!\!\right) - \left(\!\!\binom{\lceil L/2\rceil}{\lceil N/2\rceil}\!\!\right)\frac{\sigma_0(\min(L,N))}{L}. \tag{S9}$$

Employing Eqs. (S2), (S8), and (S9) we find that, as long as the right side of the above inequality is positive, the contribution of cycles with length $L_\alpha < L$ satisfy the inequality

$$\frac{\sum_{\{m_\alpha\}}\mathsf{Cyc}\left(\frac{L}{m_\alpha},\frac{N}{m_\alpha}\right)}{\mathsf{Cyc}_{\mathrm{tot}}(L,N)} < \frac{\left(\!\!\binom{\lceil L/2\rceil}{\lceil N/2\rceil}\!\!\right)\sigma_1(\min(L,N))}{\left(\!\!\binom{L}{N}\!\!\right) - \left(\!\!\binom{\lceil L/2\rceil}{\lceil N/2\rceil}\!\!\right)\sigma_0(\min(L,N))}. \tag{S10}$$

For $L \gg 1$, at unit filling ($N = L$), using Eqs. (S9), (S10) and the approximation $\left(\!\!\binom{\lceil L/2\rceil}{\lceil L/2\rceil}\!\!\right) \approx 2^{-L}\left(\!\!\binom{L}{L}\!\!\right)$, we find

$$\mathsf{Cyc}(L,L) > \left(\!\!\binom{L}{L}\!\!\right)\frac{1}{L}\left(1 - 2^{-L + \frac{1.5379 \ln(L)}{\ln\ln(L)}}\right), \tag{S11}$$

and for $L \geq 6$:

$$\frac{\sum_{\{m_\alpha\}}\mathsf{Cyc}\left(\frac{L}{m_\alpha},\frac{L}{m_\alpha}\right)}{\mathsf{Cyc}_{\mathrm{tot}}(L,L)} < \frac{Le^\gamma\left(\ln\ln L + 1\right)2^{-L}}{1 - 2^{-L + \frac{1.5379\ln(L)}{\ln\ln(L)}}}, \tag{S12}$$

where we have used known bounds on the functions $\sigma_0(n)$ and $\sigma_1(n)$ [S1, S2]. Here $\gamma = 0.57721\ldots$ is the Euler-Mascheroni constant. Thus, as reported in the main text, asymptotically, the number of cycles with length $L$ dominates, as illustrated in Fig. S1.

3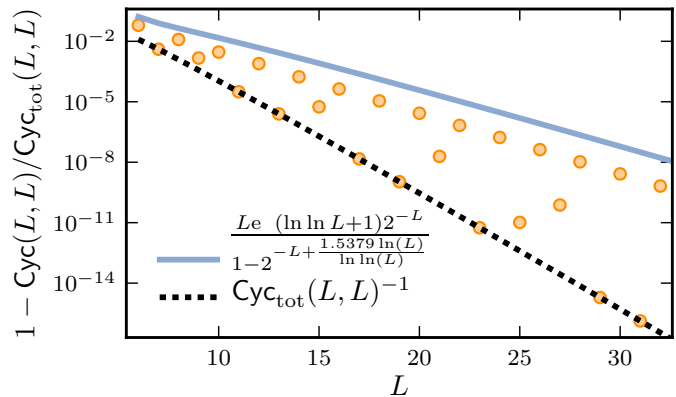

FIG. S1. The ratio of the number of cycles with length $< L$ as a function of system size $L$, at unit filling ($N = L$). The blue-solid line is the derived upper bound in Eq. (S12). The lower bound on the ratio is the inverse of the total number of cycles (black-dashed line), is satisfied if $L$ is a prime number, where all cycles have length $L$ except for one cycle that consists of the single flat occupation vector $|1, 1, \ldots, 1\rangle$.

### Inversion Symmetry

For a system with inversion symmetry, $[\widehat{H}, \widehat{R}] = 0$, where the inversion operator $\widehat{R}$ (also, known as the reflection or parity operator) acts as $\widehat{R} b_i \widehat{R}^\dagger = b_{L-i+1}$. Obviously, $\widehat{R}^2 = \mathbb{1}$, and thus the unitary operator $\widehat{R}$ is also hermitian, where $\widehat{R}^{-1} = \widehat{R}^\dagger = \widehat{R}$ such that its spectrum is $\{r = 1, r = -1\}$. Thus, we can view $\widehat{H}$ as a block diagonal matrix with two blocks, $\mathcal{M}_{r=\pm 1}$.

In general, the eigenvectors of $\widehat{R}$ with eigenvalue $r$ can be generated by pairing each occupation vector $|n_1, n_2, \ldots, n_L\rangle$ with its inversion under $\widehat{R}$ as $\frac{1}{\sqrt{2}} \left(|n_1, n_2, \ldots, n_L\rangle + r\widehat{R}|n_1, n_2, \ldots, n_L\rangle\right)$. Still, if a given occupation vector $|n_1, n_2, \ldots, n_L\rangle$ satisfies $n_i = n_{L-i+1}$, then $\widehat{R}|n_1, n_2, \ldots, n_L\rangle = |n_1, n_2, \ldots, n_L\rangle$, which could make the number of eigenvectors with $r = 1$ larger than those with $r = -1$. However, such occupation vectors cease to exist if $L$ is even and $N$ is odd. Otherwise, their number is given by $\left(\!\!\binom{\lceil L/2 \rceil}{\lfloor N/2 \rfloor}\!\!\right)$, ($\lceil \ldots \rceil$ and $\lfloor \ldots \rfloor$ are the integer ceiling and floor functions, respectively) and their contribution is exponentially small.

If the system Hamiltonian $\widehat{H}$ commutes with both of $\widehat{T}$ and $\widehat{R}$, then it is tempting to employ both translational and inversion symmetries to further simplify the diagonalization of the matrix representing $\widehat{H}$. However, in general, this cannot be achieved as $[\widehat{R}, \widehat{T}] \neq 0$.

Let's simplify $[\widehat{R}, \widehat{T}]$ by considering its action of on an arbitrary occupation vector $|n_1, n_2, \ldots, n_L\rangle$, where $\widehat{T}\,\widehat{R}|n_1, n_2, \ldots, n_{L-1}, n_L\rangle = |n_1, n_L, \ldots, n_3, n_2\rangle = \widehat{R}\widehat{T}^{-1}|n_1, n_2, \ldots, n_{L-1}, n_L\rangle$ and thus $[\widehat{R}, \widehat{T}] \equiv \widehat{R}\widehat{T} - \widehat{T}\widehat{R} \equiv \widehat{R}\left(\widehat{T} - \widehat{T}^{-1}\right)$. This implies $[\widehat{R}, \widehat{T}]|\phi_{k,\alpha,q}\rangle = 2\imath \sin(2\pi q/L)\widehat{R}|\phi_{k,\alpha,q}\rangle$, which vanishes for $q = 0$ (this is also true for $q = L/2$, when $L$ is even). Accordingly, $\widehat{R}$ and $\widehat{T}$ share a set of eigenstates in the sub-Hilbert space spanned by the basis set $\{|\phi_{k,\alpha,q=0}\rangle\}$. For $\widehat{R}|\phi_{k,\alpha,q=0}\rangle \neq |\phi_{k,\alpha,q=0}\rangle$, we have the eigenstates $\frac{1}{\sqrt{2}}\left(|\phi_{k,\alpha,q=0}\rangle \pm \widehat{R}|\phi_{k,\alpha,q=0}\rangle\right)$ with eigenvalues $q = 0, r = \pm 1$. Therefore, the matrix block $\mathcal{M}_{q=0}$ can be constructed using the two blocks $\mathcal{M}_{q=0,r=\pm 1}$ which we exploit when determining the ground state $|\Psi_0\rangle$ of the Bose-Hubbard Hamiltonian.

---


\fi

\end{document}